\documentclass[12pt]{article}

\usepackage{geometry} 

\usepackage{amsmath}
\usepackage{amsfonts}

 \usepackage{graphicx}
\usepackage{float}
\usepackage{natbib}
\usepackage{appendix}

\usepackage{hyperref}

 \usepackage[linewidth=1pt]{mdframed}
\usepackage{lipsum}
\usepackage{longtable}
 \usepackage{setspace}
\usepackage{url} 

\usepackage{sectsty}
\sectionfont{\center \fontsize{12}{15}\selectfont}
\subsectionfont{\fontsize{12}{15}\selectfont}

\newcommand\nc\newcommand
\newcommand\dmo\DeclareMathOperator
\dmo{\bin}{Bin}
\dmo{\var}{Var}
\dmo{\N}{N}
\dmo{\logit}{logit}
\dmo{\gam}{Gamma}
\dmo{\IG}{IG}
\dmo{\OR}{OR}
\dmo{\unif}{Unif}
\dmo{\UR-EE}{UR-EE}
\dmo{\NAIVE}{NAIVE}
\dmo{\MSE}{MSE}
\dmo{\LB}{LB}
\dmo{\UB}{UB}
\dmo{\B}{B}
\dmo{\auc}{auc}
\dmo{\total}{total}
\dmo{\TN}{TN}

\begin{document}

\title{Meta-Analysis of Odds Ratios With Incomplete Extracted Data}
\author{Shemra Rizzo$^{1}$ and Robert E. Weiss$^{2}$ \\
\small $^{1,2}$ Department of Biostatistics, University of California, Los Angeles, CA 90095}
\date{}

\maketitle

\begin{abstract}
A typical random effects meta-analysis of odds-ratios assumes binomially distributed numbers of events in a treatment and control group and requires the proportion of deaths to be extracted from published papers.  This data is often not available in the publications due to loss to follow-up. When the Kaplan Meier survival plot is available, it is common practice to manually measure the needed information from the plot and infer the probability of survival and then to infer a best-guess of the number of deaths. Uncertainty introduced from theses guesses is not accounted for in current models. This naive approach leads to over-certain results and potentially inaccurate conclusions.  We propose the Uncertain Reading-Estimated Events model to construct each study's contribution to the meta-analysis separately using the data available for extraction in the publications. We use real and simulated data to illustrate our methods.  Meta-analysis based on the observed number of deaths lead to biased estimates while our proposed model does not. Our results show increases in the standard deviation of the log-odds as compared to a naive meta-analysis that assumes ideal extracted data, equivalent to a reduction of the overall sample size of 43\% in our example. 
\end{abstract}


\section{INTRODUCTION}
\label{s:intro}

Meta-analysis is widely used  to quantitatively combine  results from multiple studies. In the health sciences, meta-analysis can provide stronger and more broadly-based evidence for treatment efficacy. Ideally, meta-analysis would analyze the combined individual patient data from all studies  \citep{STEWART1993, LAMBERT2002, BERLIN2002}. In practice, this is rarely done \citep{KOVALCHIK2012}. Instead, meta-analyses rely on data extracted from journal articles in the published literature or from presentations at major conferences. This is known as meta-analysis of aggregate data. 

A meta-analysis of odds ratios typically requires four quantities to be extracted per study: number of events and non-events in the treatment and control groups, which can easily be summarized in a 2x2 table. The Cochrane Handbook for Systematic Reviews of Interventions recognizes that data required for the meta-analysis are often not available in published papers \citep{COCHRANE2011}, for example, the true number of events. Software typically requires the data from the 2x2 table from each study regardless of the outcome that was computed in each study (odds, risk and hazard ratio) \citep{MELLE2004}. 
The frequent occurrence of incomplete extracted data has led to the common practice of guessing the missing entries of the 2x2 table using other information available from the study. For example, it is a very common occurrence to have all four entries missing, but the row totals are known which correspond to the numbers of people in each group at baseline. Best-guesses for the missing entries are computed from Kaplan-Meier (KM) survival curves which are often available in the published study. Because the survival curves rarely include survival probability values, meta-analysts take manual measurements from the curve to estimate them. They then multiply the survival probability and its complement by the row total of each treatment group to fill in the 2x2 table. Estimates are often rounded to the nearest integer. The guessed data is introduced in the meta-analysis as observed data, leading to unjustified certainty in the results and to potentially inaccurate conclusions. 

In some cases, the KM survival curve is not available and meta-analysts use the number of observed events reported in the text as the best approximation of the true number of events. However, in the presence of loss to follow-up this is an underestimate of the true number of events and could lead to biased meta-analysis estimates.  A typical meta-analysis includes a combination of studies where the KM probabilities are available for extraction and studies where only the observed numbers of events are available.

While there have been efforts to promote better reporting practices \citep{RILEY2003}, there is no established protocol for addressing the missing information encountered in the data extraction step of meta-analysis. For time-to-event data, some methods have been proposed to handle the missing extracted data for meta-analyses of hazard ratios but not for odds ratios  \citep{PARMAR1998, TIERNEY2007}. However, the proposed methods involve calculations that are only approximate, and do not account for the uncertainty introduced by the estimation. Methods to account for uncertainty in meta-analysis of odds ratios have been proposed for data from randomized trials that includes number of missing outcomes (non-observed events) \citep{HIGGINS2008a, HIGGINS2008b}. However, in time-to-event studies the number of missing outcomes is typically not reported.

We propose the Uncertain Reading-Estimated Events (UR-EE) Bayesian model to account for the uncertainty that arises at the data extraction step of meta-analysis. Our model formally constructs a model to properly account for the contribution of each study to the meta-analysis. Our constructions do not depend on the desired data but depend rather on the actual data available from and extracted from each published study. Data available for extraction includes the number of participants at baseline, and may include one or more of the following: the rounded survival probabilities or measurements taken off the KM plot;  confidence intervals for the KM survival probabilities;  mean, variance,  median, and quartiles  of the distributions of  follow-up times; the number of observed events, and the number of people at risk at the time of interest. Because the information available for extraction is different for each study, the extracted data from each study must be modeled individually. The UR-EE model improves the validity of results from meta-analysis by accommodating all the uncertainty in the data input to the meta-analysis. 

The paper is organized as follows. We introduce two datasets in section \ref{s:datasets}. In section \ref{s:review} we briefly review the classical and Bayesian random effects models and describe the \emph{naive} approach to manipulate incomplete extracted data to be able to use the models.  In section \ref{s:UR-EEmodel} we define the UR-EE Bayesian model for meta-analysis. Results for the two datasets from UR-EE and naive methods are given in section \ref{s:results}. The paper finishes with discussion.

\section{DATASETS}
\label{s:datasets}

\subsection{\bf Unprotected left main coronary artery stenosis data}

To exemplify the different types of extracted data and proposed methodology for a meta-analysis of odds ratios, we carefully re-evaluate a published meta-analysis that compares two treatments for unprotected left main coronary artery (ULMCA) stenosis \citep{NAIK2009}. The current gold standard treatment is coronary artery bypass grafting (CABG), which portends high morbidity.  Percutaneous coronary intervention (PCI) has emerged as a plausible alternative. It is desirable to draw a definitive assessment of both treatments. The meta-analysis performed by \citet{NAIK2009} included 10 studies with a total of 3,773 patients. The meta-analysis of mortality after 1 year presents multiple challenges in the data extraction step. In all studies the type of extractable data was the same for PCI and CABG. Table \ref{t:extracteddata} is a checklist of the components available for extraction from each study. 

\begin{table}[H]
\caption{\small Components of the extracted data from each study in the ULMCA meta-analysis}
\label{t:extracteddata}
\small
\begin{tabular}{l c c c c c c c c c c c c c c c}
\hline
\hline
Study $i$			 	&  & $n_{ij}$ &  $e_{ij}$ & $r_{ij}$ & $x_{ij}^*$ & $y_{ij}^*$ & $\kappa_{ij}^*$ & $a^*_{ij-}$ & $a^*_{ij+}$ & $m_{ij}$ & $v_{ij}^2$ &$Q1_{ij}$ & $Q2_{ij}$ &  $Q3_{ij}$ \\
\hline
Brener &  & \checkmark &   & \checkmark & \checkmark & \checkmark & & & & \checkmark & \checkmark & & & \\
Palmerini	&  & \checkmark &  & \checkmark & \checkmark & \checkmark & & & & & & \checkmark & \checkmark & \checkmark	\\
Seung &   & \checkmark  &  & \checkmark & & & \checkmark & & & & & \checkmark & \checkmark & \checkmark	\\
Wu	&  & \checkmark & &  \checkmark & & & \checkmark & & & \checkmark & & & & 	\\
Sanmartin &  & \checkmark & & \checkmark & \checkmark & \checkmark & & & & \checkmark & \checkmark & & &	\\
Buszman & & \checkmark & & \checkmark & & & \checkmark & & & \checkmark & \checkmark & & &	\\
Makikallio	&   & \checkmark  & & & \checkmark & \checkmark & & & &  \checkmark & \checkmark  & & & \\
White	& & \checkmark & & & \checkmark & \checkmark & & & & & & \checkmark & \checkmark & \checkmark	\\
Serryus	&  & \checkmark & & & & & \checkmark & \checkmark & \checkmark & \checkmark & \checkmark & & & 	\\
Chieffo   &  & \checkmark & \checkmark & & & & & & & & & & & & 	\\
\hline
\end{tabular}
Note:  $n_{ij}$ is the number of people at baseline,  $e_{ij}$ is the number of observed deaths, $r_{ij}$ is the number of people at risk at year 1, $x^*_{ij}$ and $y^*_{ij}$ are the measurements of the KM plot from the x-axis to the curve at baseline and year 1, $\kappa^*_{ij}$ is the rounded KM survival probability and ($a_{ij-}, a_{ij+}$) is its confidence interval, $m_{ij}$ and $v^2_{ij}$ are the mean and variance of the follow-up times, $Q1_{ij}$ is the median and $[Q1_{ij}, Q3_{ij}]$ is the interquartile range of the follow-up times for study $i$ and group $j$. The PCI ($j=1$) and CABG ($j=0$) groups have the same type of extractable data but different values.
\end{table}

\normalsize 

The ten studies have different types of extractable data:
\begin{enumerate}
\item All ten papers provide the number of people enrolled at baseline by treatment group.
\item Two papers provide the number of deaths observed after one year. The number of observed deaths is less than or equal to the true number of deaths, which is unknown due to loss to follow-up.
\item Four papers provide the number of people at risk after one year.
\item Eight papers provide a measure of central tendency and spread of the follow-up times. Two papers provide pooled follow-up times only. One paper provides the mean of the follow-up times by group but no variance.
\item Seven papers have a KM survival plot. Three of these plots have numerical values for the survival probabilities at year 1. For the remaining four plots, the values must be manually extracted from the plot using a ruler either in the computer screen or in print. An additional paper has a mortality rate plot with rounded mortality rates. 
\item One paper provides the observed number of deaths by treatment group and does not mention follow-up times.
\end{enumerate}

This meta-analysis motivates our development of appropriate methodology to incorporate the uncertainty that arises during data extraction into the meta-analysis model.

\subsection{\bf Simulated data}

To illustrate the dangers of using observed events as a replacement of true number of deaths in a meta-analysis we construct an extreme meta-analysis comprising ten studies, with 100 subjects on average in each arm.  The true odds ratio was set to be equal to one.  However, loss-to-follow up times were set to be considerably different: 50\% in the treatment group and  3\% in the control group by year 1. The data can be found in Appendix B. We use this data to compare the results obtained when we assume that a KM survival plot is available for all studies and when it is not available for any. 

\section{CLASSICAL AND BAYESIAN RANDOM EFFECTS META-ANALYSIS MODELS}\label{s:review}

In this section we briefly review three models for estimating the true population effect size for binomial outcomes, where the estimate of interest is an odds ratio: 1)  the classical random effects model using the popular estimates in  \citet{DERSIMONIAN1986} and the maximum likelihood (ML) estimates, and 2) the Bayesian random effects model. Then we describe the \emph{naive} approach to dealing with incomplete data to use these models.

All studies cannot be considered to be equivalent experiments. Between-study variation refers to  differences in design, execution and population, and these are reflected in the underlying true odds-ratios found in each study. The random effects model assumes that the odds ratios  in each study follow a distribution.  Within-study variation is modeled as random sampling error. Thus, the random effects model has two variance components to explain the variation in odds ratios. 

Let $n_{ij}$ be the number of subjects at baseline in study $i$, $i=1, \hdots, k$ and treatment group $j$, where $j \in\{1 = \text{treatment}, 0 = \text{control} \}$. The numbers of subjects, $s_{ij}$,  that experience the outcome event (say, death) in each group are independent binomial random variables with  $\pi_{ij}$  probability of dying before year 1, $s_{ij} | n_{ij}, \pi_{ij} \sim \bin(n_{ij}, \pi_{ij})$
for $ i\in\{1,\dots, k\}$ and  $j\in \{ 0, 1 \}$.
Study $i$'s odds ratio (OR) is $\OR_i = [\pi_{i1}/(1-\pi_{i1})]/[\pi_{i0}/(1-\pi_{i0})]^{-1}$.
Let $\mbox{O}_i=\log \OR_i$ be the observed log-odds ratio for study $i$ and $\delta_i$ be the true log-odds ratio.  The random effects model is
\begin{align}
\label{random effects1}
\mbox{O}_i | \delta_i &\sim \N(\delta_i,  \sigma^2_i),\\
\label{random effects2}
\delta_i &\sim \N(d,  \tau^2),
\end{align}
where $d$ is  the population mean of true log-odds ratios and the parameter of interest.
The between-study variation is $\tau^2$ and the within-study variance is  $\sigma_i^2$.\\

\subsection{\bf DerSimonian and Laird estimates}

In the classical random effects model, $d$ can be estimated as a weighted average of the observed log-odds ratios
\begin{equation}
\label{DSL estimate}
\hat{d} = \frac{ \sum_i w_i  \mbox{O}_i}{\sum_i w_i }
 \end{equation} 
 with variance $\widehat{\var}(\hat{d}) = 1/\sum_i w_i$,  where the weights are $w_i = [\sigma^2_i + \tau^2]^{-1}.$
In practice, the variances $\tau^2$ and $\sigma^2_i$ are unknown. Estimated variances are used instead and the effect of this practice is generally ignored \citep{BROCKWELL2001}.
The estimate of $\sigma^2_i$ is the estimated  sampling variance for an odds ratio 
\begin{equation}
\label{sampling variance}
\hat{\sigma}^2_i = \widehat{\var}[\mbox{O}_i] = \frac{1}{s_{i1}} + \frac{1}{n_{i1}-s_{i1}} + \frac{1}{s_{i2}} + \frac{1}{n_{i2}-s_{i2}}.
\end{equation}
The most widely used estimate of  $\tau^2$ is the  \citet{DERSIMONIAN1986} estimator (DSL)
\begin{equation}
\label{eq:DSLtau}
\hat{\tau}^2 = \max \left \{ 0, \frac{Q_a - (k-1)}{\sum_i a_i - \sum_i a_i^2/  \sum_i a_i}  \right \},
\end{equation}
where $a_i = 1/\hat{\sigma}^2_i$, and $Q_a= \sum_i a_i (\mbox{O}_i - \hat{d})^2$ is the Cochran statistic of heterogeneity. 

\subsection{\bf Maximum Likelihood estimates}

Likelihood estimation of $\tau^2$  is an alternative to the DSL estimator in \eqref{eq:DSLtau} \citep{VIECHTBAUER2005}. The random effects model can be written as  $\mbox{O}_i \sim \N (d, \sigma_i^2 + \tau^2)$ and the log-likelihood function is 
\begin{equation}
\label{loglikelihood}
\log L(d, \tau^2 | \mbox{O}_1, \dots, \mbox{O}_k) = \log\left(\frac{1}{(2\pi)^{k/2}}\right) -\frac{1}{2} \sum_i \log(\sigma_i^2 + \tau^2 ) - \frac{1}{2}\sum_i \frac{(\mbox{O}_i-d)^2}{\sigma^2_i + \tau^2}.
\end{equation}
Because $\sigma^2_i$ is unknown, $\sigma^2_i$ is replaced by  $\hat{\sigma}^2_i$ from  \eqref{sampling variance}. To obtain the maximum likelihood estimates we take the derivatives of \eqref{loglikelihood} with respect to $d$ and $\tau^2$ and set resulting equations to zero.  
After some manipulation, we obtain
\begin{align}
\label{maxlikeestimates2} 
\hat{\tau}^2 &= \sum_i \frac{(\mbox{O}_i - \hat{d})^2-\hat{\sigma}_i^2}{\hat{\sigma}^2_i + \hat{\tau}^2} \left[ \sum_i \frac{1}{\hat{\sigma}^2_i + \hat{\tau}^2} \right] ^{-1}. 
\end{align}
Equations  \eqref{DSL estimate} and \eqref{maxlikeestimates2} can be solved  by iterating between  $\hat{d}_t = f(\hat{\tau}^2_{t-1})$ and $\hat{\tau}^2_t = f(\hat{d}_{t}),$ with starting value $\hat{\tau}^2_0 = 0$ \citep{VIECHTBAUER2005}. 

\subsection{\bf Bayesian model}

The Bayesian random effects model allows us to include information or beliefs that may be of importance for the research question being addressed by assigning prior distributions to the parameters in the model. The fully Bayesian model accounts for all parameter uncertainty \citep{SMITH1995, CARLIN1992}.

There are several ways to perform a Bayesian random effects model for meta-analysis. Some prefer the Normal model in \eqref{random effects1} and \eqref{random effects2}  with  priors for $d$, $\sigma^2$ and $\tau^2$.
An alternative formulation preferred by some authors \citep{SMITH1995,  SUTTON2001} models the outcomes $s_{ij}$ as binomially distributed and relates the observed log-odds ratios to the probability of success using a logit transformation. Let each study's true log-odds ratio be $\delta_i = \logit(\pi_{i1}) - \logit(\pi_{i0})$, where $\logit(\pi_{ij}) =  \log (\pi_{ij}/(1-\pi_{ij} ))$, and define $u_i=[ \logit(\pi_{Ti}) + \logit(\pi_{Ci})]/2$ as the logit scale average death rate for the $i^{\mbox{th}}$ trial. 
The model is
\begin{align}
\label{eqn: bayesian model 1}
s_{ij} | \pi_{ij} &\sim \bin(\pi_{ij}, n_{ij}) \\
\label{eqn: bayesian model 2}
\logit (\pi_{i0}) &= u_i - \delta_i/2 \\
\label{eqn: bayesian model 3}
\logit (\pi_{i1}) &= u_i + \delta_i/2 \\
\label{eqn: bayesian model 4}
\delta_i &\sim \N(d, \tau^2) \\
\label{eqn: bayesian model 5}
\mu_i &\sim \N(m, \sigma^2) 
\end{align}
with priors $p(d)$, $p(\tau^2)$, $p(m)$, and $p(\sigma^2)$. By expanding this model, we are able to account for the uncertainty introduced by the data extraction.

\subsection{\bf The naive approach to incomplete extracted data}
\label{s:naive}

The classical and the Bayesian random effects meta-analysis models require four quantities to be extracted from all studies: $n_{i1}, n_{i0}, s_{i1}$ and $s_{i0}$ for all $i$. Because in most studies there is loss to follow-up, $s_{ij}$'s are unknown and \emph{best-guesses} or observed events are used instead. Meta-analysts use the KM survival probability to estimate the number of deaths. Let $\kappa^*_{ij}$ be the extracted reading of the KM survival probability in study $i$, treatment group $j$, which could be a rounded value extracted from the text or the ratio of two measurements off the KM plot. The two measurements off the plot are $x_{ij}^*$, the distance between the x-axis and the KM survival curve at year 1 and $y^*_{ij}$, the distance between the x-axis and the KM survival curve at year 0. Then, $\kappa_{ij}^* = x_{ij}^*/y_{ij}^*.$ Let  $\kappa^+_{ij}$ be the actual KM survival estimate at year 1 that is available in the computer output, and that  $\kappa^*_{ij}$ is approximating.
 
The number of deaths calculated based on $\kappa_{ij}^*$, and $\kappa^+_{ij}$  are  $s^*_{ij}=n_{ij}(1-\kappa_{ij}^*),$ and $s^+_{ij} = n_{ij} (1 - \kappa^+_{ij}),$  respectively. Estimates $s^*_{ij}$ and $s^+_{ij}$ are not necessarily integers.
The \emph{naive} approach to meta-analysis is to use $s^*_{ij}$ as $s^+_{ij}$ and, in turn, $s^+_{ij}$ as $s_{ij}$.  Then these values are fed into the classical or Bayesian random effects models. The values for $s^*_{ij}$ for the ULMCA dataset can be found in Table 2. 

 \begin{table}[ht]
 \begin{center}
 \caption{\small Extracted data for the ULMCA dataset according to the naive approach.\normalsize}
 \label{t:naivedata}
 \begin{tabular}{l  c c r  r@{.}l   c  r r@{.}l }
 \hline
 \hline
Study	&	& 		\multicolumn{4}{c}{PCI} 		&   	\multicolumn{4}{c}{CABG} 		  \\
	 	& 	&  $\kappa_{i1}^*\times100\%$     & $n_{i1}$ 	& \multicolumn{2}{c}{$s^*_{i1}$} 	& $\kappa_{i0}^*\times100\%$     & $n_{i0}$ 	&  \multicolumn{2}{c}{$s^*_{i0}$}  \\
\hline
\citet{BRENER2008}	&  	& 93.3\%		& 97		& 6&52		&  94.2\%		& 190	& 11&08   \\
\citet{PALMERINI2006} 	& 	& 89.2\%		& 154	& 16&68 		& 87.1\%		&157 	& 20&28 \\			
\citet{SEUNG2008}  	& 	& 96.7\%		& 542	& 17&89 		& 96.3\%		& 542	& 20&05 \\
\citet{WU2008}  		& 	& 83.9\%		& 135 	& 21&74 		& 94.1\%		& 135 	& 7&97 \\ 	
\citet{SANMARTIN2007} & 	& 88.5\%		& 96 		& 11&04 		& 83.5\%		& 245	& 40&55 \\
\citet{BUSZMAN2008} & 	& 98.1\%		& 52		& 0&99 		& 92.5\%		& 53		& 3&98 \\
 \citet{MAKIKALLIO2008}	& 	& 94.4\%		& 49		& 2&74 		& 89.0\%		& 238 	& 26&23 \\
\citet{WHITE2008}  	& 	& 89.8\%		& 67		& 6&83 		& 93.2\%		& 67		& 4&56 \\
\citet{SERRYUS2008} 	& 	& 95.8\%		& 357	& 14&99 		& 95.5\%		& 348 	& 15&66 \\
\citet{CHIEFFO2006} 	& 	& 	--		& 107 	& 3 & $0^a$			& 	--	& 142	& 9 & 0$^a$ \\
\hline
\hline
\end{tabular}

\small Note: $^a$ Observed number of deaths extracted from the published paper. \normalsize
\end{center}
\end{table}

Let ${\bf S}^*= ( s^*_{ij})$, ${\bf S}^+=(s^+_{ij})$, $\boldsymbol{\kappa}^* = (\kappa^*_{ij})$ and $\boldsymbol{\kappa}^+= (\kappa^+_{ij})$, $i=1,\dots, k ,j=1,0$. Let $\boldsymbol{\theta}$ be the vector of parameters $(\pi_{ij}, \delta_i, \mu_i)$ in the model. The naive Bayesian model computes the posterior 
\begin{equation}
f(\boldsymbol{\theta}| {\bf S}^*) \propto f({\bf S}^*|\boldsymbol{\theta}) f(\boldsymbol{\theta}),
\end{equation} 
and incorrectly uses $f(\boldsymbol{\theta}| {\bf S}^*)$ as a replacement for $f(\boldsymbol{\theta}| {\bf S}^+)$, ignoring that $\boldsymbol{\kappa}^*$ are approximated or rounded values of $\boldsymbol{\kappa}^+$, and in turn uses $f(\boldsymbol{\theta}| {\bf S}^+)$ as a substitute for $f(\boldsymbol{\theta}|{\bf S})$, ignoring that ${\bf S}$ are not observed but estimated from $\boldsymbol{\kappa}^+$, which are estimators themselves with additional associated uncertainty.

\section{THE UNCERTAIN READING-ESTIMATED EVENTS MODEL}\label{s:UR-EEmodel}

We propose the Uncertain Reading-Estimated Events (UR-EE) model, which does not substitute ${\bf S}^*$ for ${\bf S}^+$ for ${\bf S}$. Instead, it incorporates the uncertainty in the estimator $\boldsymbol{\kappa}^+$ by averaging over the possible values of true deaths ${\bf S}$ given $\boldsymbol{\kappa}^+$
\begin{equation}
f(\boldsymbol{\theta}|{\bf S}^+) =  \int f(\boldsymbol{\theta}|{\bf S})f({\bf S}|{\bf S}^+)d{\bf S},
\end{equation}
and over  the possible values of ${\bf S}^+$ given the extracted ${\bf S}^*$
\begin{equation}
f(\boldsymbol{\theta}|{\bf S}^*) = \int f(\boldsymbol{\theta}|{\bf S}^+)f({\bf S}^+ | {\bf S}^*)d{\bf S}^+,
\end{equation}
to obtain a posterior of the parameters given the extracted data ${\bf S}^*$
\begin{equation}\label{UR-EE posterior}
f(\boldsymbol{\theta}|{\bf S}^*) = \int \int f(\boldsymbol{\theta} | {\bf S})f({\bf S}|{\bf S}^+)f({\bf S}^+|{\bf S}^*) d{\bf S}^+ d{\bf S}.
\end{equation}
We call $f({\bf S}^+|{\bf S}^*)$ the Uncertain Reading (UR) density and $f({\bf S}|{\bf S}^+)$ the Estimated Events~(EE) density. The UR density captures the uncertainty due to not having the exact KM survival probability. The EE density captures the uncertainty of the number of deaths estimated using the KM estimator around the true number of deaths due to censoring.

Both the naive and the UR-EE model compute a $f(\boldsymbol{\theta}|{\bf S}^*)$ posterior.  The naive model is overly optimistic while the UR-EE model does not make the incorrect assumption that $f(\boldsymbol{\theta}|{\bf S}^*)=f(\boldsymbol{\theta}| {\bf S}^+)=f(\boldsymbol{\theta}|{\bf S})$.

Due to the additional incorporated  uncertainty  in the UR-EE model, we expect that 
\begin{equation}
\label{variance relationships}
\var_{\UR-EE}(d|{\bf S}^*) > \var_{\NAIVE}(d|\mathbf{S}^*).
\end{equation}
The incorporation of additional, previously ignored uncertainty in the model translates to a reduction in the effective sample size of the meta-analysis. Let $n = \sum n_{ij}$, and let $n_{\UR-EE}$ be the effective sample size of the meta-analysis under the UR-EE model. Then 
\begin{equation}
\label{sample size}
n_{\UR-EE} =  \frac{\var_{\NAIVE}(d|{\bf S}^*) }{\var_{\UR-EE}(d|{\bf S}^*) }  n,
\end{equation}
and we expect that $n_{\UR-EE} < n.$

In any Bayesian model we have observed data and parameters (random variables). In the naive Bayesian model for meta-analysis the ``observed" data is not actually observed but a guess with $s^*_{ij}$ substituting for $s_{ij}$. In contrast, the UR-EE model's observed data is the extracted data, $n_{ij}, \kappa^*_{ij}$,  and the unknown $s^+_{ij}$ and $s_{ij}$ are treated as random variables.

The fully Bayesian UR-EE model adds models 
\begin{align}
\label{eqn: UR-EE model 1}
f_{ij}(s_{ij}^*|  s_{ij}^+) \\
\label{eqn: UR-EE model 2}
f_{ij}(s_{ij}^+| s_{ij}) 
\end{align}
for all $i$ and $j$ to equations \eqref{eqn: bayesian model 1} to \eqref{eqn: bayesian model 4}. We call equation \eqref{eqn: UR-EE model 1} the Uncertain Reading density and equation \eqref{eqn: UR-EE model 2} the Estimated Events density.  The choices of  densities for  \eqref{eqn: UR-EE model 1}  and  \eqref{eqn: UR-EE model 2} are different for every $i$ due to the differences in extracted data in each study. In our example, \eqref{eqn: UR-EE model 1} and \eqref{eqn: UR-EE model 2} have the same form for $j=0, j=1$, however that is due to the reporting in the studies and not a requirement  of our methodology. We describe the construction of the UR and EE densities next.

\subsection{\bf The Uncertain Reading Density}\label{UR}

The UR density $f_{ij}(s^+_{ij}|s^*_{ij})$ is obtained from $f_{ij}(\kappa^+_{ij}|\kappa^*_{ij})$, which models the KM survival probability values based on the extracted rounded or manually measured $\kappa^*_{ij}$.  Because $\kappa_{ij}^*$ could be extracted in at least two ways, the UR density is different for each study $i$.  

\emph{Case 1: Rounded $\kappa_{ij}^*$}. In studies by Seung, Wu, Buszman and Serryus the rounded value of the survival probability $\kappa_{ij}^*$ is extracted from the text or a number printed on the KM plot. Assuming that the probability was rounded to three-digit accuracy, we model the actual KM survival probability $\kappa^+_{ij}$ as uniformly distributed  centered at $\kappa_{ij}^*$
\begin{align}
\label{kappa equation}
\kappa^+_{ij}|\kappa^*_{ij} & \sim \unif(\kappa_{ij}^*-0.0005, \kappa_{ij}^*+0.0005), \\
\label{kappa equation2}
s^+_{ij}|s^*_{ij} & \sim  \unif(s^*_{ij} - 0.0005n_{ij}, s^*_{ij}+ 0.0005n_{ij}).
\end{align}
These equations can easily accommodate rounding to different levels of accuracy by changing the minimum and maximum values in \eqref{kappa equation}-\eqref{kappa equation2}. 

\emph{Case 2: Measured $\kappa_{ij}^*$}. Let  $x_{ij}$ and $y_{ij}$ be the true unknown distances from the x-axis to the KM survival curve at year 0 and 1, and let $x^*_{ij}$ and $y^*_{ij}$ be the measurements taken off the KM plot from the x-axis to the survival curve at year 0 and year 1. Then $\kappa^*_{ij} = x^*_{ij}/y^*_{ij}$. To account for the measurement error in $x^*_{ij}$ and $y^*_{ij}$ we assume that the uncertainty in $x^*_{ij}$ and $y^*_{ij}$ is similar to rounding error, in that both are measured by a ruler with equally spaced tick marks, and that the maximal error in $x^*_{ij}-x_{ij}$ and $y^*_{ij}-y_{ij}$ is known, typically $1/2$ the distance $w_{ij}$ between the tick marks. Then $x_{ij} \sim \unif(x^*_{ij}-(w_{ij}/2), x^*_{ij}+(w_{ij}/2))$, $y_{ij} \sim \unif(y^*_{ij}-(w_{ij}/2), y^*_{ij}+(w_{ij}/2)),$ and set the true unknown KM survival probability to be $\kappa^+_{ij} = x_{ij}/y_{ij}.$
The density $f_{ij}(\kappa^+_{ij} | x_{ij}^*,  y_{ij}^*)$ is given by the ratio of two uniform random variables and has positive support on the range of values
\begin{equation}
\label{eq:support}
\frac{x^*_{ij}-\frac{w_{ij}}{2}}{y^*_{ij}+\frac{w_{ij}}{2}} \leq \kappa^+_{ij} \leq \frac{x^*_{ij}+\frac{w_{ij}}{2}}{y^*_{ij}-\frac{w_{ij}}{2}}.
\end{equation}
The exact form of the piecewise  density $f_{ij}(\kappa^+_{ij} | x_{ij}^*,  y_{ij}^*)$ is given in Appendix A. Using $f_{ij}(\kappa^+_{ij} | x_{ij}^*,  y_{ij}^*)$, an exact density $f_{ij}(s^+_{ij} | x_{ij}^*,  y_{ij}^*)$ is immediate, but we derive a distribution to be a convenient normal approximation to $f_{ij}(s^+_{ij} | x_{ij}^*,  y_{ij}^*)$ that is 
\begin{align}
\label{eq:approximation}
s^+_{ij}|s^*_{ij} &\sim \N \left (s^*_{ij}, \left (\frac{n_{ij}}{6} \right )^2\left [   \frac{x^*_{ij}+\frac{w_{ij}}{2}}{y^*_{ij}-\frac{w_{ij}}{2}} - \frac{x^*_{ij}-\frac{w_{ij}}{2}}{y^*_{ij}+\frac{w_{ij}}{2}} \right]^2 \right). 
\end{align}

This has mean and mode at $s^*_{ij}$ and standard deviation such that 99.7\% of values fall within the support in \eqref{eq:support}. Considering that the exact UR distribution, $f_{ij}(s^+_{ij}|s^*_{ij})$, is a complicated piecewise function, and that this case holds for several of the studies in the meta-analysis, a normal approximation facilitates implementation of the proposed methodology. 

\subsection{\bf The Estimated Events Density}\label{EE}

The KM survival estimate $\kappa^+_{ij}$ has two kinds of uncertainty associated with it: binomial sampling and additional uncertainty due to censoring. The binomial sampling is addressed naturally in \eqref{eqn: bayesian model 1}. 
The EE distribution addresses the additional uncertainty due to censoring. The EE distribution, $f_{ij}(s_{ij}|s^+_{ij})$, conditions on the estimated number of deaths $s^+_{ij}=n_{ij}(1-\kappa_{ij}^+)$, calculated using $\kappa^+_{ij}$ as input and gives the density of the random variable $s_{ij}$ as output. This distribution is constructed using other information from the $i^{\mbox{\tiny th}}$ paper. Since the EE distribution models numbers of deaths, a discrete distribution that assigns a probability to each feasible integer value of $s_{ij}$ is desirable. However, it is not straightforward to determine said probabilities. Thus, we approximate $f_{ij}(s_{ij}|s^+_{ij})$ with a truncated normal density,  $\TN(s^+_{ij}, \B_{ij}, \LB_{ij}, \UB_{ij})$, centered around $s^+_{ij}$ with variance $\B_{ij}$, truncated at lower and upper bounds, $\LB_{ij}$ and $\UB_{ij}$, where the values of $\B_{ij}$, $\LB_{ij}$ and $\UB_{ij}$ are dependent on the number of censored people, $c_{ij}$, in each study. 
A lower bound, $\LB_{ij}=e_{ij}$, assumes that all censored people survived and an upper bound $\UB_{ij}=e_{ij}+c_{ij}$ assumes that all censored people died. Improved lower and upper bounds can be computed with additional information available for extraction, such as number of people at risk $r_{ij}$. Simulations  (not shown) showed that asymmetry of the density $\TN(s^+_{ij}, \B_{ij}, \LB_{ij}, \UB_{ij})$ results in biased estimates of $s_{ij}$. Thus, we propose a truncated normal with symmetric truncation points 
\begin{equation}
\label{EEdistribution}
s_{ij}|s^+_{ij} \sim \TN(s^+_{ij}, \B_{ij}, s^+_{ij}-\min\{s^+_{ij}-\LB_{ij}, \UB_{ij}-s^+_{ij}\}, s^+_{ij}+\min\{s^+_{ij}-\LB_{ij}, \UB_{ij}-s^+_{ij}\}).
\end{equation}
To estimate the number of censored people, we use information extracted from the papers about the follow-up distribution times and number of people at risk found in the paper.
We model the log follow-up times as $\N(\psi_{ij}, \phi_{ij})$, with mean $\psi_{ij}$ and variance $\phi_{ij}$.  Let $\lambda_{ij}$ be the probability of being censored before year 1, then the estimated number of censored subjects is $c_{ij}=n_{ij}\lambda_{ij}$. The information about follow-up times available for extraction in each paper varies; some papers give means and variances, others give quartiles. We enumerate the following cases of extracted data to calculate $\psi_{ij}$ and $\phi_{ij}$.

\begin{description}
\item \emph{Case 1:  Follow-up time mean and variance by treatment group}. Let $m_{ij}$, $v_{ij}$  be the extracted mean and variance of the follow-up times. Then, $\psi_{ij} = \log \left [ m^2_{ij} (v_{ij} + m_{ij}^2)^{-1/2} \right ] ,$
and $\phi_{ij} = \log \left ( 1 + (v_{ij}/m^2_{ij})\right)$. 

\item \emph{Case 2: Follow-up time median and  lower and upper quartiles by treatment group}. Let $Q2_{ij}$ and  $[Q1_{ij}, Q3_{ij}]$ be the extracted median and upper and lower quartiles. We set $\psi_{ij} = \log(Q2_{ij})$ and $\phi_{ij} = \left[ \log(Q3_{ij}) - \log(Q1_{ij}) \right ] / \left[ \Phi^{-1}(0.75) - \Phi^{-1}(0.25)\right ]$.

\item \emph{Case 3: Follow-up time not by treatment group}. When the follow-up time summary statistics are pooled summaries of the treatment and control groups, set $\psi_{i1} = \psi_{i0}$ and $\phi_{i1} = \phi_{i0}$ and calculate both given the pooled summary statistics.

\item \emph{Case 4: Follow-up time mean but no variance.} We use the mean of the standard deviations in the other studies of the meta-analysis as a value for $v_{ij}$.
\end{description}

To define $\UB_{ij}$ and $\LB_{ij}$, we need the number of subjects at risk, $r_{ij}$,  and the number of observed deaths $e_{ij}$ at year 1. The availability of this information varies across studies. We consider the following cases of information available for extraction. 

\begin{description}
\item \emph{Case 1: Observed deaths and people at risk are both given in the paper}. Define $\UB_{ij} = n_{ij} - r_{ij}$ and $\LB_{ij} = e_{ij}$.

\item \emph{Case 2: Observed deaths is given but people at risk is not}. Define $\UB_{ij} = e_{ij} + c_{ij}$ and $\LB_{ij} = e_{ij}$.

\item \emph{Case 3: People at risk is given but observed deaths is not}. Define $\UB_{ij} = n_{ij} - r_{ij} $ and $\LB_{ij} =\max \{0, n_{ij} - r_{ij} - c_{ij} \}.$

\item \emph{Case 4: Neither people at risk nor observed deaths are given}. Due to the lack of information, conservative bounds are $\UB_{ij}=n_{ij}$ and $\LB_{ij}=0$.
\end{description}

The variance $\B_{ij}=\var(n_{ij}(1-\kappa_{ij}^*))=n_{ij}^2b_{ij}$ depends on, $b_{ij}$, the variance of the KM survival probability.  However, studies rarely report KM confidence intervals or KM standard errors so $b_{ij}$ is unknown. To approximate the value of the variance we simulated studies with characteristics  similar to the studies contributing to the ULMCA dataset. We found that the preferred formula depends on the amount of censoring. 
\begin{enumerate}
\item \emph{Greenwood simplified estimate}. When censoring at 12 months is less than 25\%, a simplified version of the Greenwood formula was satisfactory $b_{ij}=(\kappa^*_{ij})^2e_{ij} /[n_{ij}(n_{ij}-e_{ij})]$.
\item \emph{Censoring proportional estimate}. When censoring at 12 months ranged from 25\% to 35\%,  we found the Greenwood simplified estimate and the censoring proportional estimate, $b_{ij}=(c_{ij}/n_{ij}) \kappa^*_{ij}(1-\kappa^*_{ij})$ to be very close to each other. So we suggest the average of the two. When censoring ranged from 35\% to 50\% the censoring proportional estimate was superior while the Greenwood simplified estimate consistently underestimated the KM variance.
\item \emph{Follow-up area under the curve (AUC) proportional estimate}. The censoring proportion $c_{ij}/n_{ij}$ does not take into consideration for how long a censored subject was followed before being lost. Let $\auc_{ij}$ be the area under the curve that represents the person-years lost to follow-up, and the total area be $\total_{ij}=n_{ij}*1$ person-years, then the follow-up AUC proportional estimate is $b_{ij}= (\auc_{ij}/\total_{ij}) \kappa^*_{ij}(1-\kappa^*_{ij})$. For censoring that ranged from 50\% to 70\%, we found this estimate to be as good as the censoring proportional estimate, so we suggest using the average of the two. For censoring in excess of 70\% the AUC proportional estimate was adequate while the censoring proportional estimate consistently overestimated the KM variance.
\end{enumerate}

\subsection{\bf Survival probabilities are not available for extraction}

When a paper does not include a KM plot but it includes the number of observed deaths, the naive approach is to use the number of observed deaths as a replacement for the number of true deaths. This underestimates the true value and leads to biased information being input into the meta-analysis. 
We propose the following estimate of the probability of death at year 1, $s_{ij}/n_{ij}$ when only observed deaths are available
\begin{equation}
\label{eq:obs estimate}
k^*_{ij} = \frac{e_{ij}}{n_{ij}} \left ( \frac{1}{1-\auc_{ij}} \right),
\end{equation}
for $0< e_{ij}/n_{ij} < 0.5$ and $0<\auc_{ij}<0.5$. Then set $\kappa^+_{ij}=k^*_{ij}$. In our simulations, \eqref{eq:obs estimate} had smaller mean square error than $e_{ij}/n_{ij}$ in estimating $s_{ij}/n_{ij}$ for all ranges of censoring and true values of $s_{ij}/n_{ij}$. 

\section{RESULTS}\label{s:results}

\subsection{\bf Prior specification}
\label{s:prior}

We use proper weakly informative  priors. It is unlikely that the underlying odds ratio associated with the model would exceed 100 in favor of either PCI or CABG \citep{SMITH1995}.  With this constraint, a range for $d$ is ($-4.6$, $4.6$). Thus, a normal  $\text{N}(0, 2.35^2)$ prior for $d$ has the mean $\pm$ 1.96 SD covering 95\% of the prior range.
The event (death) rate $\logit^{-1}$ is very likely to lie in the interval $(0.001, 0.999)$ \citep{LAROSE1997}, which corresponds to ($-3.89$, $3.89$) as an interval for $m$. Assuming a normal distribution, the prior is $\text{N}(0, 1.98^2)$.
We use informative Inverse Gamma priors $\IG(3,2)$ for the variance parameters $\sigma^2$ and $\tau^2$ \citep{GELMAN2004}. 

\subsection{\bf Conditional densities}
\label{ss:densities}

Figure 1 compares the conditional densities $f(\pi_{ij} | s^*_{ij})$ using the naive Bayesian model (dashed lines) and $f(\pi_{ij} | \text{extracted data})$ using the UR-EE model (solid lines) in the ULMCA meta-analysis of mortality at year 1. The dotted line corresponds to the use of the ratio of two uniform random variables as described in Appendix A, while the dashed line uses the normal approximation in \eqref{eq:approximation}.  Thus,  equation \eqref{eq:approximation} serves as an adequate approximation for the ratio of two uniform random variables, in our situation. Curves corresponding to the Chieffo paper are not included as the extracted data did not include follow-up information. Sensitivity analysis (not shown) showed that multiple assumptions about censoring, including no censoring, in Chieffo's data did not change the ULMCA meta-analysis result. 

\begin{figure}[H]
\centering
\caption{\small Naive conditional densities, $f(\pi_{ij} | s^*_{ij})$ are in solid lines and UR-EE conditional densities $f(\pi_{ij} | \text{extracted data})$ are in dashed lines. Where applicable, the dotted lines (in red) uses the ratio of two uniform random variables while the dashed line (in blue) uses the normal approximation.}
\includegraphics[scale=0.48]{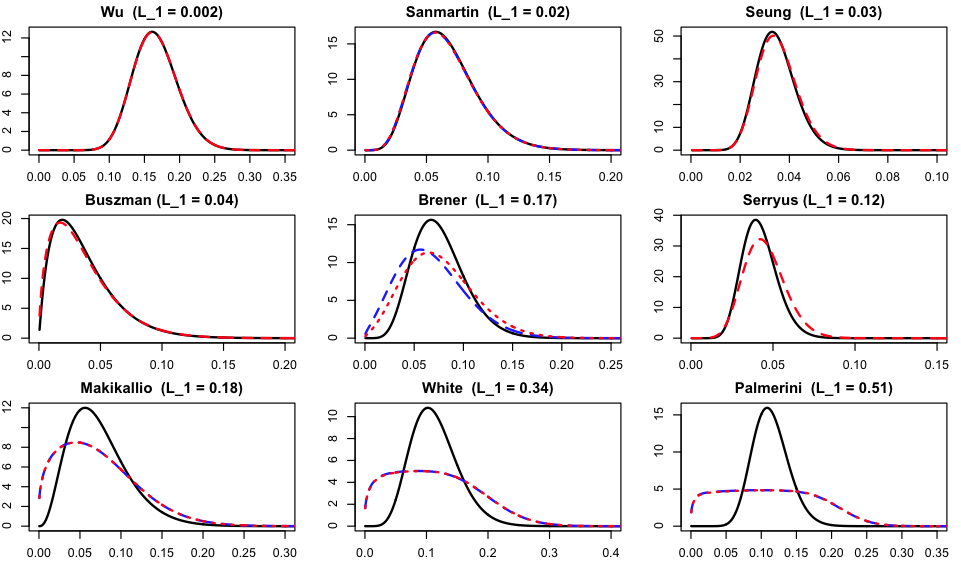}
\end{figure}

After accounting for the uncertainty that arises from the estimates of the number of deaths, there is an increase in the variance of all conditional densities, confirming the conservativeness of our model. To quantify the effect, we use the $L_1$  statistic to measure the distance between the naive and proposed densities \citep{WEISS1996}. The statistic takes values [0,1], where a value of zero indicates no difference and a value of one indicates maximal difference between the densities. The densities in Figure 1 are sorted in increasing order of $L_1$.  The effect of accounting for the different types of  extracted data varies greatly among studies.  Minimal differences between the naive and proposed densities are found for the studies by Wu, Sanmartin,  Seung  and Buszman where $L_1 < 0.05$.  Large differences of $L_1 > 0.2$ are found in the studies by White and Palmerini.  The densities suggest that having a smaller amount of extractable data available from a study results in large increases in the study OR's variance compared to having full information.

\subsection{\bf Posterior Computation}

Because the posterior is not available in closed form, we base our inferences on an MCMC simulation from the posterior distribution $f(\boldsymbol{\theta}| {\bf S}^*)$.  We use a Gibbs step \citep{GELFAND1990} whenever conditional posteriors are available, as for $d$, $\sigma^2$, $m$, $\tau^2$. When the conditional distribution is of intractable form, such as for $m_i$ and $\delta_i$, we use a Metropolis step \citep{METROPOLIS1953}, resulting in a hybrid sampler \citep{TIERNEY1994}. Gibbs steps are included to sample for $s^+_{ij}$, and $s_{ij}^*$ using the Uncertain Reading and Estimated Events densities. 

The MCMC algorithm to compute the posterior of the UR-EE model was implemented in R \citep{R2008} using the ``\verb'ureepkg'.'' The package was developed specifically for the ULMCA and simulated data meta-analyses, and is available upon request from the authors. After an initial burn-in period of 2,000 iterations, we generated an additional $100,000$ iterations, retaining every tenth iteration. We used three chains with different starting points and assessed convergence by inspecting the densities and time series plots and the convergence diagnostics of \citet{GELMAN1992}. 

 \subsection{\bf Sensitivity to prior specification}
 \label{ss:sensitivity}
  
We examine sensitivity of the inference of $d$ to the prior specification of the variance components. We use $\IG(\epsilon, \epsilon)$ as an attempt at uninformativeness \citep{GELMAN2006} for the prior distributions of $\sigma^2$ and $\tau^2$.  The resulting estimates for the log-odds ratio and its standard deviation do not vary substantially for different choices of prior when using $\epsilon \in \{ 0.1, 1\}$ when compared to those obtained using an $\IG(3,2)$ as a prior.  
  
\subsection{\bf Results with ULMCA dataset}
\label{ss:comparison}
  
Table \ref{table:bayesian results} summarizes the results from the naive and UR-EE Bayesian models or the meta-analyses of mortality of the ULMCA dataset. The results  confirm that there is an increase in variance in all parameters as expected from the discussion around \eqref{variance relationships}. The increase in the standard deviation of the log-odds ratio is 33\%.  Across all parameters in the model, there was an average increase of 39\% in the standard deviation. 

\begin{table}[H]
\caption{\small Posterior mean and 95\% credible intervals for parameters $d$, $\sigma^2$, $m$ and $\tau^2$ in the naive and UR-EE Bayesian  models for mortality of the ULMCA dataset}\label{table:bayesian results}
\centering
\begin{tabular}{l l c c c c }
\hline
\hline
\multicolumn{2}{l}{}  &  \multicolumn{2}{c}{Mean}   &	 \multicolumn{2}{c}{SD}	\\
 &				& Naive  	& UR-EE   	&  Naive 	& 	UR-EE  \\
\hline
Moratilty\\
&$d$		 	& -0.09		& -0.07		 & 0.22 	&	0.29 	 \\
&$\sigma^2$ 		& 0.25		& 0.47		 &   0.29	& 	0.48	\\
&$m$			&-2.65		& -2.71		 &   0.18	& 	 0.20	\\
&$\tau^2 $ 		& 0.23		& 0.29		&   0.	17	& 	0.24 	\\
\hline
\hline
\end{tabular}
\end{table}
  
Figure 2 gives $\OR$ estimates and 95\% intervals for the mortality meta-analysis under the DSL, ML, naive Bayes and UR-EE models. As expected the UR-EE model interval is wider than the naive models. According to \eqref{sample size}, the increase in the standard deviation over the naive Bayes model is equivalent to a reduction of 43\% of the meta-analysis sample size. 

\begin{figure}[H]
\centering
\caption{\small Morality meta-analysis for ULMCA dataset: odds-ratio mean (Bayes and UR-EE) or point estimate (DSL, ML), $\exp(d)$, and 95\% confidence interval under various models.}
\includegraphics[scale=0.55]{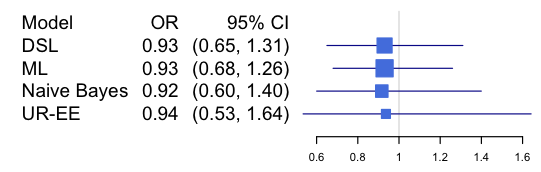}
\end{figure}

\subsection{\bf Results with simulated dataset}

The simulated dataset has large censoring (50\%) in the treatment arm and little to no censoring (3\%) among controls at year 1 in all ten studies. 
Log follow-up times $\N(\psi_{ij}, \phi_{ij})$ were set to be $N(5.89, 0.83)$ and $\N(7.05, 0.63)$, for all $i$, in the treatment and control groups respectively. The follow-up times are measured in days. The values for the parameters in \eqref{eqn: bayesian model 2} to \eqref{eqn: bayesian model 5} were set to $d=1.0$, $\tau^2=0.4$, $m=-0.8$, and $\sigma^2=0.1$.  
Figure 3 shows the results obtained under different scenarios. Because this is simulated data we are able to compute the odds ratio and confidence interval using the true number of deaths, which is the ideal extracted data that is typically unavailable.  If we assume that the KM survival probabilities were available for extraction in both arms in all studies then we can compute naive Bayes and UR-EE models to obtain similar conclusions as in the ULMCA dataset. The UR-EE confidence interval is wider than the one obtained under the naive Bayes model, and both are close to the one obtained from the model based on the true data. 
The power of the UR-EE model is portrayed in the extreme case where KM survival probabilities are not available in all studies. The meta-analysis that uses observed deaths as a substitute for true deaths results in a biased and inaccurate result: it points to a significant difference in two arms where none exists! Using the UR-EE model on observed deaths results in an adequate odds ratio and confidence interval. 

\begin{figure}[H]
\centering
\caption{\small Mortality meta-analysis simulated: odds-ratio mean $\exp(d)$ and 95\% confidence interval under various models.}
\includegraphics[scale=0.55]{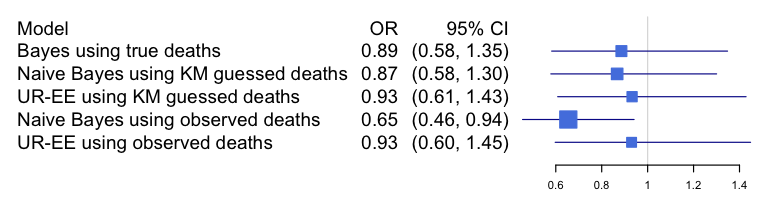}
\end{figure}

\section{DISCUSSION}\label{discussion}

In the ULMCA meta-analysis we re-evaluated, the estimates from ML were slightly more conservative than those obtained using DSL.  The naive Bayesian model was more conservative than both DSL and ML, and the UR-EE model was, as expected, the most conservative of all.  The odds ratio for mortality reported in \citet{NAIK2009}, OR=1.00  [95\% CI: 0.70 to 1.41], did not use any of the naive methods. Instead it incorporated uncertainty in the unknown number of deaths by reducing the sample sizes of each study. Naik's interval is wider than the ones from DSL and ML but not as wide as the Bayesian naive and UR-EE models.

The UR-EE model penalizes the lack of important information available for extraction in the papers by increasing the variance in the estimates of the study level and global parameters. Given that each study can present its own challenges, our model potentially requires individual models carefully constructed for each study's contribution to the meta-analysis. While the algorithm increases in complexity the more studies it contains, the results are more conservative and more accurate than traditional meta-analysis methods such as DSL, ML or a naive Bayesian model, which do not account for the uncertainty of their extracted data. 

Under the UR-EE model, there is an increase in the standard deviation of all parameters in the model. Thus, non-significant results would be even more non-significant. The conclusion for the ULMCA meta-analysis was the same regardless of the method: there is no statistical difference in the rates of mortality between PCI and CABG.  While the UR-EE model did not change the conclusions here, in other datasets where the odds ratio intervals are different but close to 1.0, it is possible that under the UR-EE model, the wider interval would include 1.0 causing the results to become non-significant. For example, if in our meta-analysis of mortality, the estimated log-odds $\hat{d}$ had been within $(-0.57, -0.43)$ or $(0.43, 0.57)$  instead of zero, with the same standard deviation as in Table \ref{table:bayesian results}, the resulting confidence intervals for the odds ratio would be significant in the naive model and non-significant in the UR-EE model. 

Our model's superiority over the naive models was best exemplified in the extreme case of the simulated data. The data was simulated such that a naive approach would result in biased and inaccurate results, which is the case where observed values are used instead of true values. When feeding that same data into our UR-EE model, we obtained accurate results. While it was also shown that a naive meta-analysis based on best-guesses obtained from KM survival probabilities was adequate in our situation, a typical meta-analysis contains a mix of studies where KM plots are available and where they are not and observed values are used instead, making our UR-EE model a necessary requirement in the computation of meta-analyses of odds-ratio with incomplete extracted data. 

We have presented a model that incorporates the uncertainty that arises during data-extraction in the meta-analysis model. Our model would not be necessary if published studies provided better estimates and standard errors and complete information on their follow-up times. Unfortunately, the poor reporting of summary statistics will continue to prevail in the published literature, making the UR-EE model a requirement for any meta-analysis performed that does not include individual patient data. While the overall conclusion may not change from those obtained from naive approaches, our method allows researchers to explore the impact of the uncertainty from the missing extracted data in the final estimates. For example, if a meta-analysis yields a significant result using standard methods, we strongly recommend our method be run in parallel to confirm the significance of the results after accounting for the uncertainty of the missing extracted data to avoid the situation portrayed in our simulated dataset.

To improve estimates from meta-analyses, we recommend that referees and editors of journals require submitted papers to include complete survival plots that contain KM estimates and numbers of people at standardized time points,  confidence bands for the KM estimates for both treatment groups,  and more complete follow-up time information for both groups, with minimum means, standard deviations and quartiles and maximum times. If these data are not going be part of the published paper, it should be made available in online supplemental materials for future extraction.

\begin{appendix}
 
\section{Appendix A: Ratio of two uniformly distributed random variables.}
Let $x^{\text{true}} \sim \text{Unif}(x-w, x+w)$ and  $y^{\text{true}} \sim \text{Unif}(y-z, y+z)$, where $x>w>0$ and $y>z>0$. Then, the density of $ p = x^{\text{true}}/ y^{\text{true}}$ is

for $\frac{x-w}{y-z} \leq \frac{x+w}{y+z} $
\begin{displaymath}
g(p | x, y, w, z)   =  \left \{ \begin{array}{lcl}
0 & \text{for} & p  \leq \frac{(x-w)}{(y+z)} \\
 \frac{(y+z)^2 - \frac{(x-w)^2}{(p)^2}}{8wz} & \text{for} &  \frac{(x-w)}{(y+z)} \leq p   \leq \frac{(x-w)}{(y-z)} \\
 \frac{y}{2w} & \text{for} &  \frac{(x-w)}{(y-z)} \leq p  \leq \frac{(x+w)}{(y+z)} \\
 \frac{\frac{(x+w)^2}{(p)^2} - (y-z)^2}{8wz} & \text{for} &  \frac{(x+w)}{(y+z)} \leq p \leq \frac{(x+w)}{(y-z)} \\
0 & \text{for} & \frac{(x+w)}{(y-z)} \leq p,   \\
\end{array} \right.
\end{displaymath}

for $\frac{x-w}{y-z} \geq \frac{x+w}{y+z} $
  \begin{displaymath}
g(p | x, y, w, z)   =  \left \{ \begin{array}{lcl}
0  & \text{for} & p  \leq \frac{(x-w)}{(y+z)} \\
 \frac{\frac{(x-w)^2}{(p)^2} - (y+z)^2}{8wz}  & \text{for} &  \frac{(x-w)}{(y+z)} \leq p  \leq \frac{(x+w)}{(y+z)} \\
 \frac{x}{2z(v)^2}  & \text{for} &  \frac{(x+w)}{(y+z)} \leq p  \leq \frac{(x-w)}{(y-z)}\\
 \frac{\frac{(x+w)^2}{(p)^2} - (y-z)^2}{8wz} & \text{for} & \frac{(x-w)}{(y-z)} \leq p  \leq \frac{(x+w)}{(y-z)}\\
0 & \text{for} & \frac{(x+w)}{(y-z)} \leq p . \\
\end{array} \right.
 \end{displaymath}

\section{Appendix B: Simulated data}

\vspace{-5mm}

 \begin{table}[H]
 \begin{center}
 \caption{Simulated data}
 \label{t:simulateddata1}
 \begin{tabular}{c c c c  c c c c c c c}
 \hline
 \hline
Study	& $n_{i1}$ & $e_{i1}$ 	&   $s_{i1}$  	& $\kappa_{i1}$ & lost to  		& $n_{i0}$ & $e_{i0}$ &   $s_{i0}$  	& $\kappa_{i0}$ & lost to\\
		&		&		  	&			& 			& follow-up  	&		&		  &			& 			& follow-up\\
		&		&			&			&			&  $j=1$		&		&		   &			&			&  $j=0$	\\
\hline
 1  		&   45    	&         17      	&        23    	&      0.53  		& 16    		&  32        	&      12     	&          12   	&       0.63   	&            0       	\\
 2 		&   139   	&         21     	&         31    	&      0.82   	&  60    		&   36        &       6      	&          6       	&      0.83    	&          0     	\\
 3 		&   75      	&       25        	&      30       	&    0.59    	 	&    29   		&  15         &      4        &        4         	&    0.73  		&             0     	\\
 4 		&    80     	&        31        	&      35       	&   0.53  		&     22   		&    23       &       10   	&          10      	&      0.57	   	&           0    	\\
 5 		&  156     	&        43      	&        55      	&     0.66 		&    66     		&  187       &       72   	&         73       	&     0.61    	&          4      	\\
 6 		&    87      	&       23      	&        30      	&    0.68  		&     35     		&      90     &        19  	&          20       	&     0.79     	&          5    	\\
 7		&  183     	&       21       	&       24     	&    0.87  		&   72      		&   116      &       32   	&         32       	&     0.72   		&             3  	\\
 8		&  141      	&       26       	&       29    	  	&    0.78  		&    62    		&    70       &     22     	&       22        	&    0.69   		&           0    	\\
 9		& 158       	&      27        	&      32     	 	&    0.80   		&    68    		&     168    &        63   	&         64      	&      0.62    	&         2        	\\
10		 & 120      	&       38       	&       45     	 &    0.60  		&      47   		&     48      &       20   	&       20        	&    0.58      	&      1      		\\
\hline
\hline
\end{tabular}
\end{center}
\small Note:  $n_{ij}$ is the number of people at baseline,  $e_{ij}$ is the number of observed deaths, $s_{ij}$ is the number of true deaths and $\kappa_{ij}$ is the KM survival probability at year 1, for study $i$ and group $j$. \normalsize
\end{table}

\end{appendix}

\vspace{5mm}

\bibliographystyle{apalike}
\bibliography{REFERENCES}

\end{document}